\documentclass{article}

\usepackage{microtype}
\usepackage{graphicx}
\usepackage{subfigure}
\usepackage{booktabs} 

\usepackage{hyperref}

\usepackage[accepted]{icml2024}

\usepackage{amsmath}
\usepackage{amssymb}
\usepackage{mathtools}
\usepackage{amsthm}

\usepackage[capitalize,noabbrev]{cleveref}

\theoremstyle{plain}

\theoremstyle{definition}

\theoremstyle{remark}

\usepackage[textsize=tiny]{todonotes}

\icmltitlerunning{A solution for the mean parametrization of the von Mises-Fisher distribution}

\input m.symbols

\author{%
  Marcel Nonnenmacher\\ 
  Gatsby Computational Neuroscience Unit \\ 
  University College London \\
  London, UK.\\
  \texttt{m.nonnenmacher@ucl.ac.uk} \\
   \And
  Maneesh Sahani \\
  Gatsby Computational Neuroscience Unit \\ 
  University College London \\
  London, UK.\\
  \texttt{m.sahani@ucl.ac.uk}
}

\begin{document}

\twocolumn[
\icmltitle{A solution for the mean parametrization of the von Mises-Fisher distribution}
\begin{icmlauthorlist}
\end{icmlauthorlist}

\begin{center}
 {\bf Marcel Nonnenmacher \hspace{0.5cm} Maneesh Sahani} \\
 \vspace{0.2cm}
    Gatsby Computational Neuroscience Unit, University College London, London, UK
\end{center}

\icmlcorrespondingauthor{Marcel Nonnenmacher}{m.nonnenmacher@ucl.ac.uk}
\icmlcorrespondingauthor{Maneesh Sahani}{m.sahani@ucl.ac.uk}

\icmlkeywords{Computational Statistics, Machine Learning}

\vskip 0.3in
]


\begin{abstract}
  The von Mises-Fisher distribution as an exponential family can be expressed in terms of either its natural or its mean parameters. Unfortunately, however, the normalization function for the distribution in terms of its mean parameters is not available in closed form, limiting the practicality of the mean parametrization and complicating maximum-likelihood estimation more generally.  
  We derive a second-order ordinary differential equation, the solution to which yields the mean-parameter normalizer along with its first two derivatives, as well as the variance function of the family. We also provide closed-form approximations to the solution of the differential equation. This allows rapid evaluation of both densities and natural parameters in terms of mean parameters. 
  We show applications to topic modeling with mixtures of von Mises-Fisher distributions using Bregman Clustering. 

\end{abstract}


\section{Introduction}

The von Mises-Fisher (vMF) distribution is a classical distribution for data located on the unit hypersphere.  It is an important tool in directional statistics \citep{mardia1975statistics} and emerges naturally when considering data in the form of normalised vectors or embeddings \citep{banerjee2005clusteringhyper} as is common in machine learning.
It is the maximum entropy distribution given the first moment $\mathbb{E}[X]$ of a variable defined on the unit hypersphere $X\in S^{D-1}$ \citep{rao1973linear}, and has been used for clustering of angular data \citep{banerjee2005clusteringhyper}, and more recently as a latent variable distribution in variational autoencoder frameworks \citep{davidson2018hyperspherical, xu2018spherical}.

In exponential family form, the vMF distribution can be described by the density with respect to the uniform measure $\nu$ on $S^{D-1}$,
\begin{align}    
p(x | \eta) = e^{\eta^\top{}x - \Phi(\eta)}, \nonumber
\end{align}
i.e. with 
natural parameter $\eta \in \mathbb{R}^D$ and log-partition function 
\begin{align*}
    \Phi(\eta) &= \log \int e^{\eta^\top{}x} \nu(dx)  \nonumber \\ &=\log I_{\frac{D}{2}-1}(||\eta||) - \left(\frac{D}{2}-1\right) \log ||\eta|| - \frac{D}{2} \log 2\pi, \nonumber
\end{align*}
where $||\cdot||$ is the Euclidean norm and $I_v$ denotes the modified Bessel function of the first kind and order $v$.
As a natural exponential family distribution with strictly convex log-partition function, the vMF distribution can equivalently be parametrized by the expectation $\mu = \mathbb{E}[X] = \grad{}\Phi(\eta)$, also known as the mean parameter\footnote{A common notation for the von Mises-Fisher parameters is $(\mu,\kappa) \in S^{D-1} \times \mathbb{R}_{\geq{}0}$ with direction $\mu$ and concentration parameter $\kappa$. Here, we reserve $\mu$ with $||\mu||\leq 1$ to refer to the mean parameter in exponential family form, giving direction and concentration parameters of $\eta / ||\eta|| = \mu / ||\mu||$ and $\kappa=||\eta||=||\grad{}\Psi(\mu)||$, respectively.}, in the form 
\[
p(x | \mu) = e^{\grad{}\Psi(\mu)^\top(x - \mu) + \Psi(\mu)},
\]
where 
\[\Psi(\mu) = \Phi^*(\mu) = \max_\eta \mu^\top{}\eta - \Phi(\eta)\] 
is the Legendre transform $\Psi=\Phi^*$ of the log-partition function $\Phi$. 
Because $H[X|\mu] = - \mathbb{E}_{p(x|\mu)}\left[\log p(x|\mu) \right] = - \Psi(\mu)$, we will refer to $\Psi$ as the ``negative entropy'' function.
$\Psi$ is strictly convex
and of Legendre-type \citep{rockafellar1997convex}, i.e. $\lim_{||\mu||\rightarrow{}1} ||\grad\Psi(\mu)|| = \infty$ as $\mu$ approaches the domain boundary of $\Psi$ at $||\mu|| = 1$.

The mean-parametrized form has several nice properties such as a well-interpretable domain for $\Psi$ \citep{wainwright2008graphical}, and is advantageous for learning as can be seen with a simple example: assume we have independently generated angular data $x^n \in S^{D-1}, n=1,\ldots,N$, and want to fit a vMF distribution to it. 
In mean parametrization, the log-likelihood is given by
\begin{align}
\mbox{LL}(\mu) &= \frac{1}{N} \sum_{n=1}^N \log p(x^n | \mu) \nonumber \\ &= - D_\psi\left(\frac{1}{N}\sum_{n=1}^N x^n \ || \ \mu\right) + \frac{1}{N} \sum_{n=1}^N \Psi(x^n)\nonumber
\end{align}
with non-negative Bregman divergence 
\[
D_\psi(x ||\mu) = \grad\Psi(\mu)^\top(\mu-x) - \Psi(\mu) + \Psi(x) \geq 0.
\] 
Because $D_\psi(x ||\mu)=0$ if and only if $x=\mu$, maximum likelihood for mean-parametrized vMF  distributions (and exponential families in general) coincides with simple moment matching 
\[\mu_{ML} = \argmax_\mu \mbox{LL}(\mu) = \frac{1}{N} \sum_{n=1}^N x^n. \]
To evaluate the learned density $p(x|\mu_{ML})$, however, we need either to evaluate $-D_\Psi(x||\mu) + \Psi(x)$ and hence both $\Psi(\mu)$ and $\grad{}\Psi(\mu)$, or alternatively to transform mean parameters $\mu_{ML}$ into natural parameters $\eta_{ML} = \grad\Psi(\mu_{ML})$ and evaluate $p(x|\eta_{ML})$. 

Unfortunately, $\Psi=\Phi^*$ and its gradient $\grad{}\Psi(\mu)$ are not available in closed form.
Since $\grad\Psi = \left(\grad\Phi\right)^{-1}$ by the properties of the Legendre transform \citep{bauschke1997legendre}, practical maximum likelihood for vMF distributions often depends on numerical root finders for $||\grad{}\Phi(\eta) - \mu_{ML}||$ 
to obtain maximum-likelihood estimates $\eta_{ML}$.

In the following we derive a second-order ordinary differential equation whose solution under appropriate boundary conditions gives both $\Psi(\mu)$ and $\grad\Psi(\mu)$ at a precision limited only by the numerical ODE solver. We furthermore give simple approximations to $\grad\Psi(\mu)$ and $\Psi(\mu)$ based on rational functions and their antiderivatives.
We provide code in Python\footnote{See \url{https://github.com/mnonnenm/vonMisesFisher_negativeEntropy}}.

\subsection{Related work}

The duality of natural and mean parametrizations for exponential families was explored in depth by \citet{banerjee2005clustering}, who presented a general expectation maximization \citep{dempster1977maximum} algorithm for mixture models with exponential family mixture components in mean parametrization. This EM variant, dubbed `Bregman clustering' due to reliance on $D_\Psi(x||\mu)-\Psi(x)$, was however not applicable to vMF mixture components because the vMF negative entropy $\Psi$ and associated Bregman divergence were not available in closed form. 

For the special case of vMF mixture components, \citet{banerjee2005clusteringhyper} in the same year instead presented an adjusted EM algorithm operating in natural parameter space, and applied it to document clustering problems in high-dimensional feature spaces. 
This EM algorithm for mixtures of vMF (`MovMF') distributions included an approximation to the mapping from (weighted) average data norms $|| \frac{1}{N} \sum_n x^n|| \approx ||\mu||$ to vMF concentration parameters $\kappa = ||\eta||$ as needed to find the maximum likelihood natural parameters, thus approximating a scalar mapping $\psi'(||\mu||) = ||\eta||$. They discussed further refinements to their approximation, but did not discuss consequences for estimating $\Psi(\mu)$ or $D_\Psi(x||\mu)$, which would have made it possible to also use Bregman clustering in the MovMF case. 
Since then, additional studies have developed methods to numerically evaluate the mapping $||\mu||\mapsto ||\eta||$ using iterative solvers \citep{tanabe2007parameter, sra2012short, song2012high, hornik2014maximum, christie2015efficient,   hasnat2016model}.

A separate line of work in theoretical statistics investigates variance functions $V(\mu)$ to characterize natural exponential families \citep{morris1983natural, letac1990natural, bar1994diagonal, ghribi2015characteristic}. Variance functions describe the (co-)variance $\mbox{Cov}[X]$ of a random variable $X$ in terms of its mean $\mathbb{E}[X] = \mu$, and thus are related to negative entropies via the inverse Hessian, $V(\mu) = \left(\grad^2\Psi(\mu)\right)^{-1}$. To the best of our understanding, the variance function of the vMF distribution has not thus far been characterized. As we will show, it can be computed numerically by solving a second-order univariate ODE.

\section{Materials \& methods}

We derive a second-order ODE to describe $\psi''(||\mu||)$, the second derivative of the scalar-valued radial profile $\psi(||\mu||)$ of $\Psi$.  Note that the negative entropy inherits radial symmetry from $\Phi(\eta) = \phi(||\eta||)$:
\begin{align}
    \Psi(\mu) &= \max_\eta \mu^\top \eta - \Phi(\eta) = \max_\eta \mu^\top \eta - \phi(||\eta||) \nonumber \\ &= \max_{||\eta||} ||\mu|| \cdot ||\eta|| - \phi(||\eta||) \nonumber \\ &= \psi(||\mu||). \nonumber
\end{align}
The radial profile $\psi : [0, 1[ \rightarrow \mathbb{R}_{\geq{}0}$ is sufficient to also describe the gradients $\grad\Psi(\mu)$, since their direction is always aligned with $\mu$. Similarly, the Hessian $\grad^2 \Psi(\mu) \in \mathbb{R}^{D\times{}D}$ can be expressed in terms of $\psi'(||\mu||)$ and $\psi''(||\mu||)$ (see appendix \ref{appendix:radial_functions_grad_hess_det}).

\subsection{Covariances and support on the unit hypersphere}

For natural exponential families with associated log-partition function $\Phi$ and negative entropy $\Psi$, the inverse function theorem applied to pairs $(\mu,\eta)$ with $\mu = \grad\Phi(\eta)$ gives 
\[\grad^2\Psi(\mu) = \left(\grad^2\Phi(\eta)\right)^{-1} = \left(\mbox{Cov}[X|\mu]\right)^{-1}.\]

We can derive an important property of the vMF variance function $V(\mu) = \mbox{Cov}[X|\mu]$ simply from the fact that the support is restricted to $S^{D-1}$:
Any random variable $X$ with support limited to the unit ball $||X|| \leq 1$ has an upper bound on its variances and hence on the eigenvalues of $\mbox{Cov}[X]$ in terms of the mean.
For random variables with $\mathbb{E}[X] =\mu$ and support limited to the unit ball, we have (see appendix \ref{appendix:trace_ineq}) 
\[tr\left(\mbox{Cov}[X|\mu] \right) \leq 1 - \mu^\top\mu.\]
Furthermore, equality is obtained if and only if all the probability mass lies on the surface of unit ball (i.e. on the hypersphere),
\[tr\left(\mbox{Cov}[X|\mu] \right) = 1 - \mu^\top\mu \ \  \mbox{ iff } \ ||X||=1 \ \mbox{ a.s.}\]
Note that this latter result is based only on the support. It does not depend on the distribution of mass over $S^{D-1}$.

\subsection{Radial symmetry and a second-order ODE for the von Mises-Fisher negative entropy } 
\label{sec:ODE}
From the eigendecomposition $\grad^2\Psi(\mu) = U \Lambda U^\top$, we have
\[tr\left(\grad^2\Psi(\mu)^{-1}\right) = \sum_i \Lambda_{ii}^{-1} = \sum_i \frac{1}{\lambda_i}\] 
for eigenvalues $\lambda_i(\mu) > 0$ of $\grad^2\Psi(\mu)$.
For radially symmetric $\Psi(\mu) = \psi(||\mu||)$, we have (see appendix \ref{appendix:radial_functions_grad_hess_det})
\[| \grad^2\Psi(\mu) | = \left(\frac{ \psi'(||\mu||) }{||\mu||}\right)^{D-1} \psi''(||\mu||),\] 
where $|\cdot|$ on the left-hand side denotes the determinant.
The eigenvalues of the Hessian are $\lambda_1 = \psi''(||\mu||)$ (with eigenvector $\mu$) and $\lambda_{i} = \frac{ \psi'(||\mu||) }{||\mu||}, i = 2,\ldots,D$. 

Thus for the von Mises-Fisher distribution $p(x|\mu)$ with support over $S^{D-1}$ and radially symmetric $\Psi(\mu)$, we find 
\begin{align} 
tr\left(\mbox{Cov}[X|\mu]\right) &= \sum_i \lambda_i^{-1} \nonumber \\ &= \frac{1}{\psi''(||\mu||)} + (D-1) \frac{||\mu||}{\psi'(||\mu||) } = 1-||\mu||^2. \nonumber 
\end{align}
Reordering terms thus yields a second-order nonlinear ODE 
\begin{align}
    \psi''(||\mu||)   = \frac{\psi'(||\mu||)}{(1 - ||\mu||^2) \psi'(||\mu||) + (1-D) ||\mu||}. \nonumber
\end{align}
This result is exact, in that no approximations were used to arrive at this identity.

\subsection{Initial conditions}
The vMF distribution with $\eta=0$ has $\mu=\mathbb{E}_\nu[X]=0$, and by Legendre duality, $\grad{}\Psi(0) = 0$.  Thus we have 
\begin{align}
\psi(0) &= \lim_{||\eta||\rightarrow 0} -\Phi(||\eta||) \nonumber \\ &= \left(\frac{D}{2}-1\right) \log 2 + \frac{D}{2} \log 2\pi + \log \Gamma(D/2). \nonumber
\end{align}
However, setting $\psi'(0) = ||\grad{}\Psi(0)|| = 0$ in the ODE for $||\mu|| = 0$ yields an ill-defined initial condition. In practice we approximate $\psi'(0) = \lim_{\eps\downarrow{}0} \eps$ with a small value ($\leq 10^{-4}$, see appendix \ref{appendix:initial_values} for an evaluation of this). Knowing the value $\psi'(||\mu||)$ for any $||\mu|| > 0$ with sufficient precision would allow us to circumvent this issue by starting the integration from that point.

\subsection{Connection to variance functions}

Variance functions $V(\mu)$ have been recognized as a useful tool to (uniquely) characterize natural exponential families \citep{morris1983natural}. 
The vMF variance function for given radial profiles $\psi', \psi''$ is (see appendix \ref{appendix:radial_functions_grad_hess_det})
\[
V(\mu) = \frac{||\mu||}{\psi'(||\mu||)} I_D + \left( \frac{1}{\psi''(||\mu||)} - \frac{||\mu||}{\psi'(||\mu||)} \right) \frac{\mu \mu^\top}{||\mu||^2},
\]
the trace of which recovers the result in section \ref{sec:ODE}. 
Traces of variance functions have been studied as potential identifiers of natural exponential families \citep{arab2015trace}.
Due to the occurrence of $\psi', \psi''$, this description of the variance function is not closed-form.

We note that the general form of this variance function, a scaled identity matrix plus a scaled rank-one matrix $\mu\mu^\top$, also holds for other radially symmetric negative entropies with domain $||\mu||\leq{}1$, but whose associated exponential families unlike the vMF have support within the interior of the unit ball $||X||\leq{}1$, and thus for which $tr(V(\mu)) < 1-\mu^\top\mu$ for at least some $\mu$. 

We further note that computing the negative entropy $\Psi(\mu)$ from (the reciprocal of) the variance function $(V(\mu))^{-1} = \grad^2\Psi(\mu)$ involves finding a second anti-derivative. 
Even if the variance function were known in closed form, analytic anti-derivatives may not be available. Thus a numerical approach might remain necessary even in this case, e.g. through iterated line integrals starting from a point $\mu_0$ with known $\Psi(\mu_0), \grad\Psi(\mu_0)$ -- for $\mu_0 = 0$, this becomes very similar to the numerical solution of our second-order ODE.  However, the scalar form of our ODE comes from being based on a matrix trace, rather than from line integrals.

\subsection{Closed-form approximation}
\label{sec:closedform_approx}

\citet{banerjee2005clusteringhyper} gave an excellent approximation to the derivative of the negative entropy profile
\[\psi_B'(||\mu||) := \frac{||\mu|| (D-||\mu||^2)} {1-||\mu||^2} \approx \psi'(||\mu||) \]
based on a first-order truncation of a continued fraction series representation of $\phi'(||\eta||)$ \citep{watson1995treatise}, with an additive error correction term derived from post-hoc inspection.  The result is useful for a wide range of dimensionalities $D$.
Their main use was to approximate the mean-to-natural parameter mapping $\eta = \grad{}\Psi(\mu) =\frac{\psi'(||\mu||)}{||\mu||} \mu$ that occurs in maximum likelihood and expectation maximization for exponential families in natural parametrization. 
A question when wanting to work with the mean parametrization throughout is if the anti-derivative 
\[\psi_B(||\mu||) = \frac{1}{2}||\mu||^2 + \frac{1-D}{2} \log \left(1-||\mu||^2\right) + const. \]
can serve as a basis to approximate $\psi(||\mu||) = \Psi(\mu)$. 

As we will show, this anti-derivative is indeed a plausible approximation to the numerical solutions for $\psi$ obtained from our ODE approach. 
Based on this finding, and with the second-order ODE at hand, we propose an augmented numerical approximation obtained by refining $\psi_B$:
\begin{align}
    \tilde{\psi}_1'(||\mu||) &= \frac{(D-1) ||\mu||}{1- ||\mu||^2 - 1/\psi_B''(||\mu||)} \nonumber \\ &=
\frac{(1-D)\left(||\mu||^5 + (D-3) ||\mu||^3 + D ||\mu||\right)}{||\mu||^6 + (D-3) ||\mu||^4 + ||\mu||^2 + D - 1},  \nonumber 
\end{align}
which has a closed-form anti-derivative 
\begin{align}    
    \tilde{\psi}_1(||\mu||) &= \frac{1-D}{4s}\log\frac{v + ||\mu||^2 + s}{v + ||\mu||^2 - s} \nonumber \\  &+ \frac{1-D}{2} \log(1-||\mu||^2) + const. \nonumber
\end{align}
for $v=\frac{D}{2}-1$ and $s= \sqrt{\left(\frac{D}{2}-1\right)^2 -D}$.
The additive constant can easily be identified through $\tilde{\psi}_1(0) = \psi(0)$.
A further refinement can be achieved with 
\begin{align}
    \tilde{\psi}'_2(||\mu||) &=  \frac{(D-1) ||\mu||}{1- ||\mu||^2 - 1/\tilde{\psi}_1''(||\mu||) }, \nonumber
\end{align}
yielding a rational function of degree ten (see appendix \ref{appendix:diff_ODE}). Computing the antiderivative of $\tilde{\psi}'_2$ involves solving a fifth-order polynomial with coefficients given by $D$, making it somewhat cumbersome. We hence use $\tilde{\psi}_1$ in the following.

Since $\tilde{\psi}_r'', r\in\{1,2\}$ are also available in closed form, we can reformulate the ODE to numerically solve for $\psi-\tilde{\psi}_r$ and $\psi' -\tilde{\psi}_r'$ (see \ref{appendix:diff_ODE}), or for large $D\gg10$ omit numerical integration of the ODE entirely and simply approximate $\psi \approx \tilde{\psi}_1$, $\psi' \approx \tilde{\psi}_2'$.

\section{Results}

To assess the accuracy of $\psi(||\mu||) = \Psi(\mu)$ and $\psi'(||\mu||) = ||\grad\Psi(\mu)||$ obtained by numerical integration of the second-order ODE we adopted the following procedure.

We began by choosing a set of natural parameters $\eta$ of different magnitudes.  These were used to evaluate negative entropies 
\[
  -H[X|\eta] = ||\eta|| \phi'(||\eta||) - \phi(||\eta||) - \frac{D}{2} \log{}2\pi,
\] 
and the corresponding mean parameters $\mu$ by numerical calculation  of the roots of 
\[
  \|\grad\Phi(\eta)\| -||\mu|| = \phi'(||\eta||) - ||\mu|| = \frac{I_{D/2}(||\eta||)}{I_{D/2-1}(||\eta||)}  - ||\mu||
\]
with the Newton-Raphson method (see appendix \ref{appendix:newton} for details).
This procedure yielded triples $(\mu, \grad\Psi(\mu)=\eta, \Psi(\mu)=-H[X|\eta])$ based on prior knowledge of $\eta$.  We then asked whether we could recover $\Psi$ and its gradient starting from $\mu$ alone.

In fact, this process is complicated by the function $\phi'(||\eta||)$ requiring to evaluate the modified Bessel function for orders $D/2$, and $D/2-1$ when $D$ can be  large. Fast and asymptotically valid approximations for large $||\eta||$ have been known for long \citep{abramowitz1968handbook}, but in many applications with high dimensional data, we need to evaluate both narrow and broad vMF distributions. 
Several previous studies on parameter inference in vMF distributions found it necessary to develop their own approximations of $\log I_v$ and $I_v/I_{v-1}$ needed to compute the log-partition function $\Phi(\eta)$ and its gradient \citep{tanabe2007parameter, sra2012short, song2012high, hornik2014maximum}. 
Here, we truncated a continuous fraction representation of $I_{v}/I_{v-1}$ (see appendix \ref{appendix:newton}) to obtain an approximation of the ratio of modified Bessel functions.
For $\log I_v$, we followed recent work on vFM modeling \citep{kim2021pytorch} and used a library for arbitrary-precision floating point arithmetic \citep{johansson2013mpmath}, which we found sufficient for the dimensionalities $D$ explored here.

\begin{figure*}[h!]
    \centering
\includegraphics[width=0.8\textwidth]{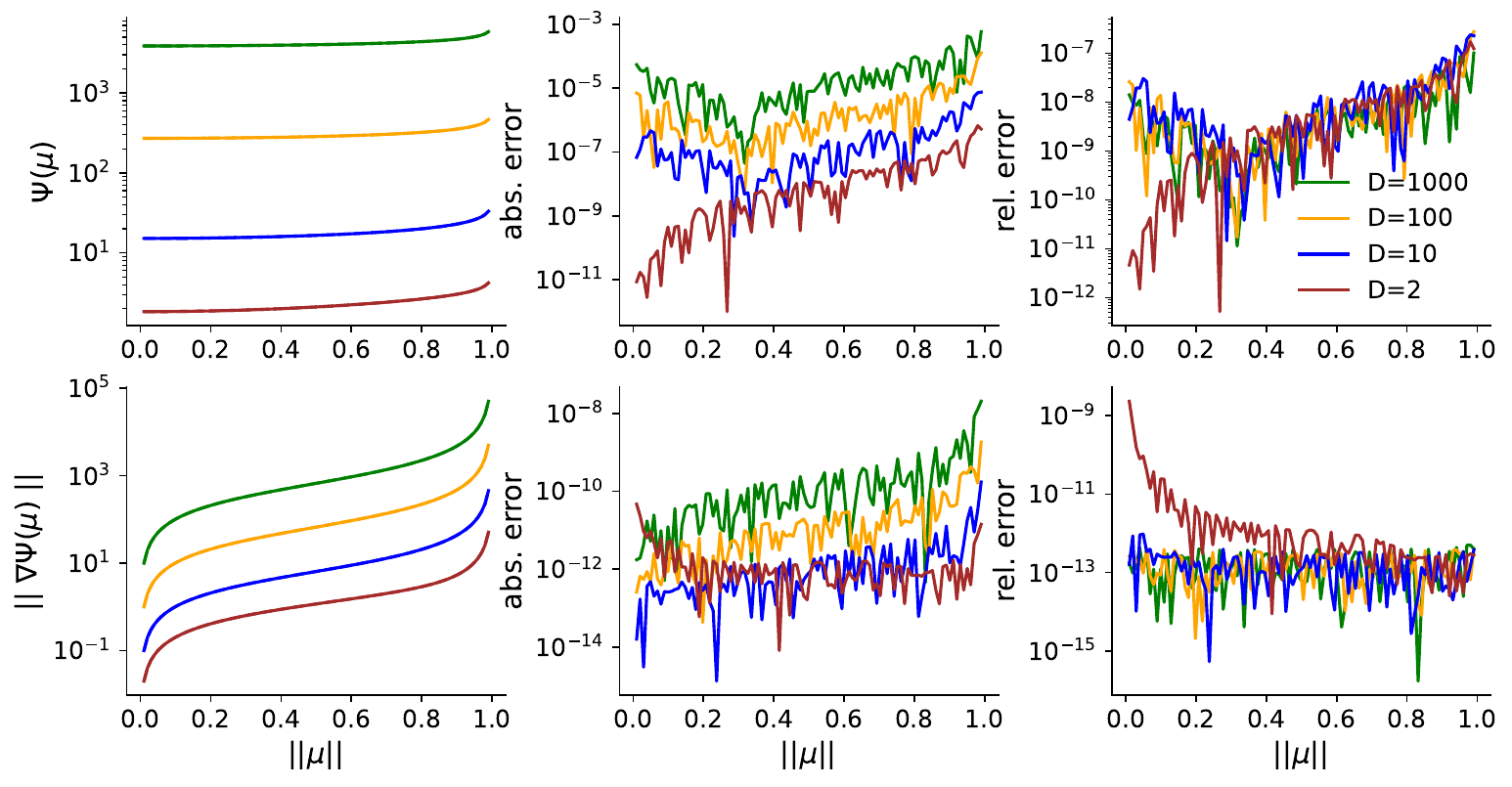}
    \caption{Numerical comparison of ODE results for the negative entropy of the vMF. 
    Top: comparison of negative entropy computed from second-order ODE and von Mises Fisher entropy $H[p|\eta]$ computed in natural parametrization. Note that the shown errors include those from mapping between natural- and mean-parameter spaces.
    Bottom: gradient of the negative entropy. Since the direction of the gradient is given by $\mu$, we only need to identify $||\grad\Psi(\mu)||=\psi'(\mu)$. We compare the numerical solution of the second-order ODE against root-finding results (Newton-Raphson on $\phi'(||\eta||) - ||\mu||$).}
    \label{fig:fig1}
\end{figure*}

When evaluating the negative entropy and its gradient at several points $\mu_k$, we sorted them by norm and solve the ODE from $||\mu||=0$ to $||\mu||=\argmax_k ||\mu_{k}||$, forcing the solver to stop at all intermediate $||\mu_k||$. This way we only needed to solve the ODE once.  
To speed up subsequent evaluations after a first solve, one can also store checkpoints of the ODE solutions at intermediate $||\mu||$ for use as initial conditions.
We used the standard Runge-Kutta 4(5) numerical solver for initial value problems from the SciPy package \citep{2020SciPy-NMeth} to numerically integrate this second-order ODE from specified initial conditions, with absolute and relative error tolerances of $10^{-12}$.
As seen in Fig.~\ref{fig:fig1}, the ODE solutions agree well with the $\eta$-derived values of both the negative entropy and its gradient.  

\begin{figure*}[h!]
    \centering
    \includegraphics[width=0.8\textwidth]{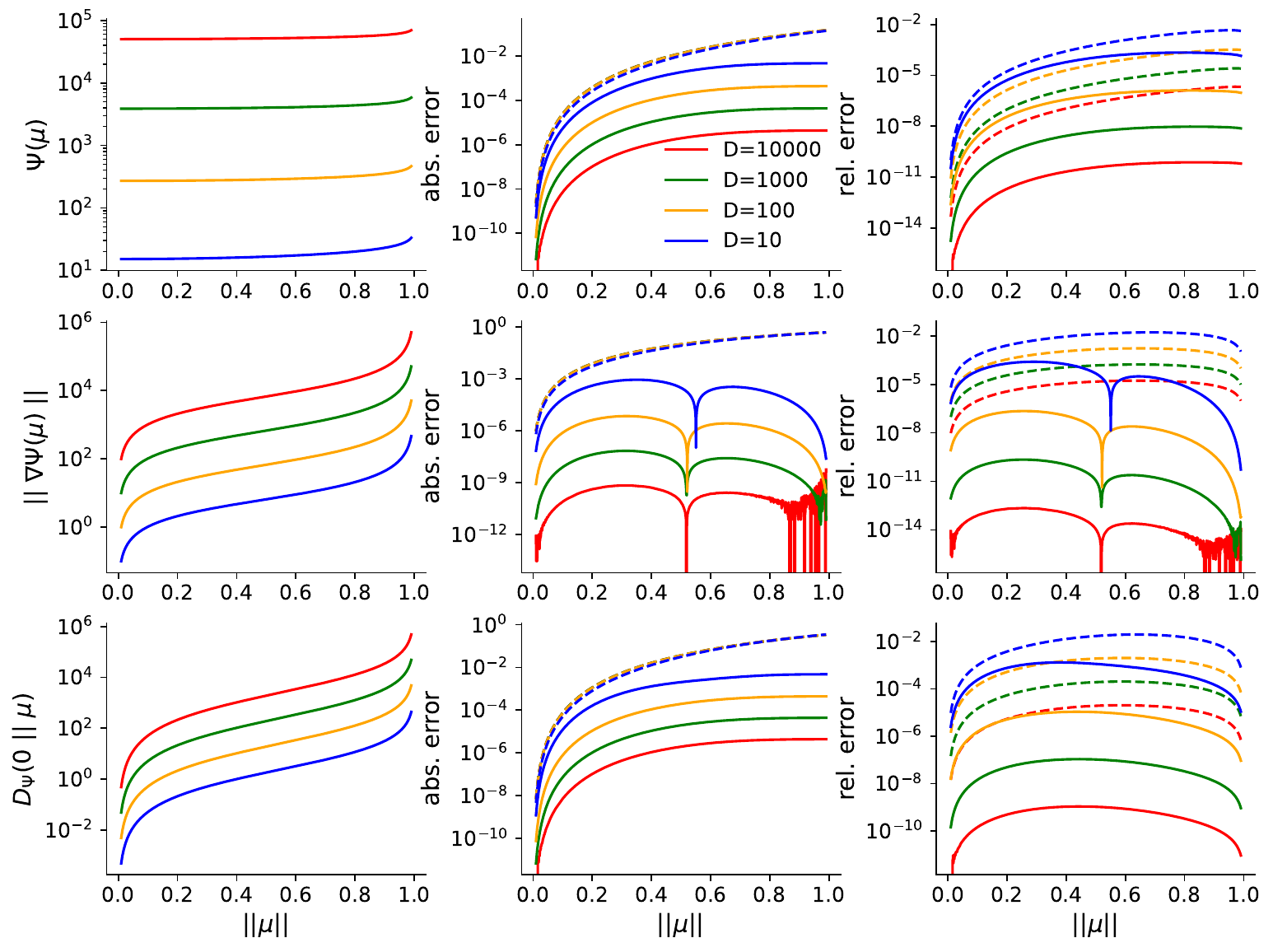}
    \caption{Evaluation of closed-form approximations for $\Psi(\mu)=\psi(||\mu||)$, $||\grad\Psi(\mu)|| = \psi'(||\mu||)$, and $D_\Psi(0 || \mu)= ||\mu|| \psi'(||\mu||)  - \psi(||\mu||)$, against numerical ODE solution, for different dimensions $D$. Left: numerical solutions to the second-order ODE in dimension $D = 10^i$, $i \in \{1,2,3,4\}$. Graphs for the numerical approximations are visually indistinguishable. Center: absolute errors between ODE solution and closed-form approximations. Right: relative errors between ODE solution and closed-form approximations. Dashed lines for approximations $\hat{\Psi}_B$ based on Banerjee et al. (2005), solid lines for our adjusted approximations $\tilde{\Psi}$ (see text). Dents in relative errors for the gradient norms come from the closed-form approximations switching between under- and overestimation.}
    \label{fig:fig2}
\end{figure*}

We next compared the various closed-form approximations to $\psi, \psi'$ (\cref{sec:closedform_approx}) and the resulting estimate of $D_\Psi(0||\mu)= \grad\Psi(\mu)^\top{}\mu - \Psi(\mu)$ against numerical solutions of the ODE (Fig.~\ref{fig:fig2}).  We computed both the absolute error magnitude $|f(\mu) - \hat{f}(\mu)|$ and the relative error $|f(\mu) - \hat{f}(\mu)| / f(\mu)$ for different values of $\mu$ and $D$, where $f \in \{\psi, \psi', D_\Psi(0 || \cdot) \}$ and $\hat{f}$ is the corresponding estimate based on $\psi_B$ or $\tilde{\psi}_r$.

While the approximation $\psi_B$ given by Banerjee et al. (dashed lines) is quite accurate, both the absolute and (to a lesser degree) relative errors tend to increase with $||\mu||$.  By contrast, the absolute error appears to depend only weakly on dimension $D\in\{10,100,1000,10000\}$, even though the values of all three of the target functions increase by several orders of magnitude from $D=10$ to $D=10^4$.  As a result, the relative errors fell with $D$. 
Our adjusted closed-form estimate $\tilde{\psi}_1$ and our estimate $\tilde{\psi}_2'$ for the derivative (solid lines) achieve lower absolute and relative errors than $\psi_B$ across the full range of investigated $||\mu||$ (Fig. \ref{fig:fig2}). Interestingly, also the absolute errors for our approximations shrink with increasing dimensionality $D$. For $D\geq 50000$, we noticed little difference in absolute error to the ODE solution between the approximation $\tilde{\psi}_2'(||\mu||)=\kappa$ and root-finding on $\phi'(\kappa) - ||\mu||$ via Newton-Raphson and common numerical approximations to $\phi' = \frac{I_{D/2}}{I_{D/2-1}}$ (see appendix \ref{appendix:newton}).


\subsection{Bregman clustering}

\begin{table}[b]
\caption{Datasets for document clustering. Number of categories $K_{true}$, number of documents $N$ and feature space dimension $D$.}
\label{sample-table}
\vskip 0.15in
\begin{center}
\begin{small}
\begin{sc}
\begin{tabular}{lcccr}
\toprule
Data set & $K_{true}$ &$N$ &  $D$ \\
\midrule
classic3    & 3 & 3891 & 4674   \\
class300    & 3 &  300 &  2551  \\
news20    & 20 & 18803 & 28571   \\
news20-small   & 20 & 2000 & 16218  \\
\bottomrule
\end{tabular}
\end{sc}
\end{small}
\end{center}
\vskip -0.1in
\label{table:datasets} 
\end{table}

Bregman clustering \citep{banerjee2005clustering} is based on expectation maximization for a mixture exponential family model given by
\[
p(x \ | \ \theta=\{\pi_k, \mu_k\}_{k=1}^K) = \sum_{k=1}^K \pi_k \ p(x \ | \ \mu_k), 
\]
with $\pi_k \geq 0, \sum_k \pi_k =1$ and exponential family $p(x|\mu)$ defined in mean parametrization. Fitting proceeds by expectation maximization, using $\log p(x|\mu) = - D_\Psi(x|\mu) + f(x)$ for efficient likelihood evaluation based on the negative entropy $\Psi$. 
Here, we exploited our novel formulation of the vMF negative entropy and derivatives to apply Bregman clustering to the problems of clustering documents \citep{banerjee2005clusteringhyper} from the 'news20' and 'classic3' datasets \citep{dhillon2003information,lang1995newsweeder}, which each comprise several thousand documents from multiple categories. 
Previous models of these data have relied on natural parameter vMF formulations, as methods to evaluate $\Psi$ were unavailable. 

Documents indexed $n=1,\ldots, N$ were mapped onto feature representations $x^n \in S^{D-1}$ using the term-frequency inverse document frequency \citep{robertson2004understanding} representation.
$D$ is given by the size of the dataset-derived vocabulary that defines the word-frequency embedding space. 
The vocabularies are large, and so the resulting representation vectors are high-dimensional with $D$ ranging in the thousands to tens of thousands. 
Following \citet{banerjee2005clusteringhyper}, we also created smaller datasets by subsampling $100$ documents per category. 
See appendix \ref{sec:datasets} for further information on datasets and feature spaces, and Table \ref{table:datasets} for a summary of dataset features.

\begin{figure}[t]
    \centering
    \includegraphics[width=0.5\textwidth]{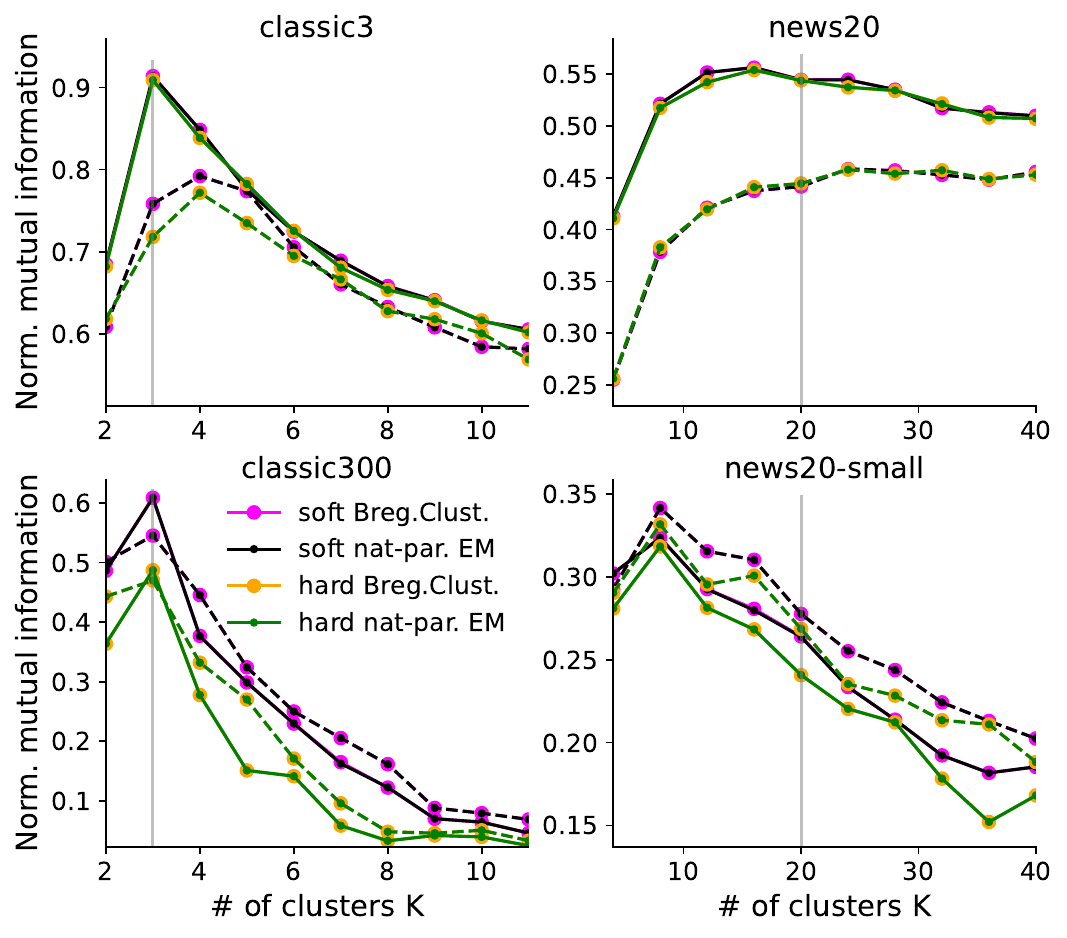}
    \caption{Clustering on the unit hypersphere. Normalized mutual information scores between true and estimated clusterings for two different variants of the document collections 'classic3' and '20newsgroup'. Documents are mapped by term-frequency inverse document frequency onto unit hyperspheres in dimensions $D=4674$ for 'classic3' and $D=28571$ for 'news20'.  Gray lines mark true number of clusters. 'classic300' and 'news20-small' are subsampled datasets (see text). We compare several EM algorithms using mixture of vMF models. Results are averaged over $10$ random seeds. Solid lines are for tied cluster concentration parameters, i.e. $||\eta_k|| = ||\eta_l||$ resp. $||\mu_k|| = ||\mu_l||$ for all $k,l=1,\ldots, K$. Dashed lines are for separate concentration parameters. We find no qualitative difference between the results of the EM algorithms working in natural parameter space (black \& green), and the Bregman clustering algorithms (magenta \& orange) that work in mean parametrization using our numerical approximations to the Bregman divergence. }
    \label{fig:fig3}
\end{figure}

We fitted MovMF models with different numbers of mixture components and algorithmic variants to each dataset. Algorithmic variants were soft- and hard-assignment EM. Soft assignment EM computes the full cluster ``responsibilities'' $p(k|x, \theta)$ in the E-step, whereas in the hard-assignment E-step we assign each datapoint $x^n$ to a single mixture component $k_n' = \argmax_k  p(k|x^n, \theta)$.
For both soft- and hard assignment variants, we fitted a MovFM model in natural parametrization $\theta=(\pi, \{\eta_k\}_k)$ and another one in mean parametrization $\theta=(\pi, \{\mu_k\}_k)$, from the equivalent initialization $\mu_k^{\mbox{\small{}init}} =\grad\Phi(\eta_k^{\mbox{\small{}init}})$ using $100$ EM iterations. We compared results across $10$ different random initializations. Bregman clustering fits used the closed-form approximation to $\psi, \psi'$ described in section \ref{sec:closedform_approx}.
Since it was previously noted  that MovMF models applied to document clustering perform better with tied concentration parameters $||\eta_k|| = ||\eta_l||$, resp. $||\mu_k||=||\mu_l||$, for all $k,l=1,\ldots,K$ \citep{zhong2003comparative} , we also included such restricted model variants (Fig. \ref{fig:fig3}, dashed lines).

We evaluated results using normalized mutual information between the true document topics and the clustering obtained from the responsibilities $\argmax_k p(k|x^n, \theta_{ML})$. 
We did not find qualitative differences in the results between the corresponding mean- and natural-parameter models and algorithms, showcasing that Bregman clustering on high-dimensional hyperspheres is feasible with our fast approximations to the negative entropy and its gradient.

\section{Discussion}

We derived a second-order differential equation to describe the negative entropy of the von Mises-Fisher distribution and its associated variance function.
To arrive at this result, we combined two features of the vMF distribution: that the negative entropy is radially symmetric, and that it only has support on the boundary of the mean-parameter domain, i.e. the unit hypersphere.

Many other radially symmetric negative entropy functions with domains $||\mu||\leq{}1$ may exist and describe valid natural exponential families over the unit ball. These distributions would be described by different radial profiles $p(||x||)$, but share the general form for the variance function of a scaled identity matrix plus rank one matrix. Conversely, other negative entropies with $tr(V(\mu))=1-||\mu||^2$ could be found which are not radially symmetric, but still describe natural exponential families with support over the unit hypersphere.
Conditions on matrix-valued functions to be valid second derivatives of strictly convex functions have been investigated in machine learning \citep{richter2021input} and statistics \citep{ghribi2015characteristic}, and could be combined with the trace condition to describe new natural exponential families on the unit hypersphere. 


We furthermore described fast approximations to the negative entropy and its first two derivatives that over a wide range of dimensionalities become better with higher dimensions. Importantly, these closed-form approximations are numerically cheap compared to iterative schemes based on numerical approximations to the vMF log-partition. In particular for small dimensions it however remains an open question how best to use the derived relations among derivatives of the negative entropy, without having to numerically solve the ODE at high precision.


\section*{Acknowledgments}
This work was funded by the Gatsby Charitable Foundation (GAT 3850) and the Simons Foundation (SCGB 543039). We thank Samo Hromadka and Florent Draye for helpful discussions.


\clearpage
\bibliographystyle{icml2024}
\bibliography{refs}


\clearpage
\appendix  \section{Appendix}   

\subsection{Gradient and Hessian of radially symmetric functions}
\label{appendix:radial_functions_grad_hess_det}
Let $f(x) = g(||x||)$ for $x \in \mathbb{R}^D$, $r=||x||\geq 0$ and $I_D$ denote the $D\times{}D$ identity matrix. Then from basic calculus and the Sherman–Morrison formula,  
\begin{align}
    (\grad{}f)(x) &= \frac{g'(r)}{r} x, \nonumber \\
(\grad^2{}f)(x) &= \frac{g'(r)}{r} I_D + \left( \frac{g''(r)}{r^2} - \frac{g'(r)}{r^3} \right) x x^\top, \nonumber \\
(\grad^2{}f)^{-1}(x) &= \frac{r}{g'(r)} I_D + \left( \frac{1}{g''(r)} - \frac{r}{ g'(r)} \right) \frac{x x^\top}{r^2}. \nonumber 
\end{align}

For radially symmetric function $f(x)$ we furthermore have 
\begin{align}
    |\grad^2{}f| 
    &= \left|\grad^2{}f\right|  \nonumber \\
    &= \left| \frac{g'(r)}{r} I_D + \left( \frac{g''(r)}{r^2} - \frac{g'(r)}{r^3} \right) x x^\top \right| \nonumber \\
    &\overset{*}{=} \left| \frac{g'(r)}{r} I_D \right| \left( 1 + \frac{r}{g'(r)} \left( \frac{g''(r)}{r^2} - \frac{g'(r)}{r^3} \right) x^\top I_D  x \right) \nonumber \\
    &= \left( \frac{g'(r)}{r} \right)^D \left( \frac{g''(r)}{r^2} \frac{r^3}{g'(r)} \right) \nonumber \\
    &= \left( \frac{g'(r)}{r}\right)^{D-1} \ g''(r),  \nonumber
\end{align}
where the identity labelled by $^*$ derives from the matrix determinant lemma. 

The eigenvalues indeed are $g''(r)$ for eigenvector $x$,
\[
\left(\grad^2{}f(x)\right) \ x = \frac{g'(r)}{r} x + \left(\frac{g''(r)}{r^2} - \frac{g'(r)}{r^3}\right) r^2 x = g''(r) x,
\]
and $g'(r)/r$ with multiplicity $D-1$ for eigenvectors $y$ from the orthogonal space with $x^\top{}y=0$,
\[
\left(\grad^2{}f(x)\right) \ y = \frac{g'(r)}{r} y.
\]

\subsection{Bound on covariance trace}
\label{appendix:trace_ineq}

Let $X$ be a $D-$dimensional random vector $X$ with mean $\mathbb{E}[X] = \mu$ and restricted support such that $||X||\leq 1$ a.s. Then
\begin{align}
tr\left( \mbox{Cov}[X] \right) &= \sum_{d=1}^D \mathbb{E}[X_d^2] - \mathbb{E}[X_d]^2 \nonumber \\  &= \mathbb{E}[\ ||X||^2\ ] - \mu^\top{}\mu \nonumber \\  &\leq 1 - \mu^\top{}\mu, \nonumber
\end{align}
with equality if and only if $\mathbb{E}[\ ||X||^2 ] = 1$.

\subsection{Dependence of ODE on exact initial conditions}
\label{appendix:initial_values}

\begin{figure}[h!]
    \centering
    \includegraphics[width=0.50\textwidth]{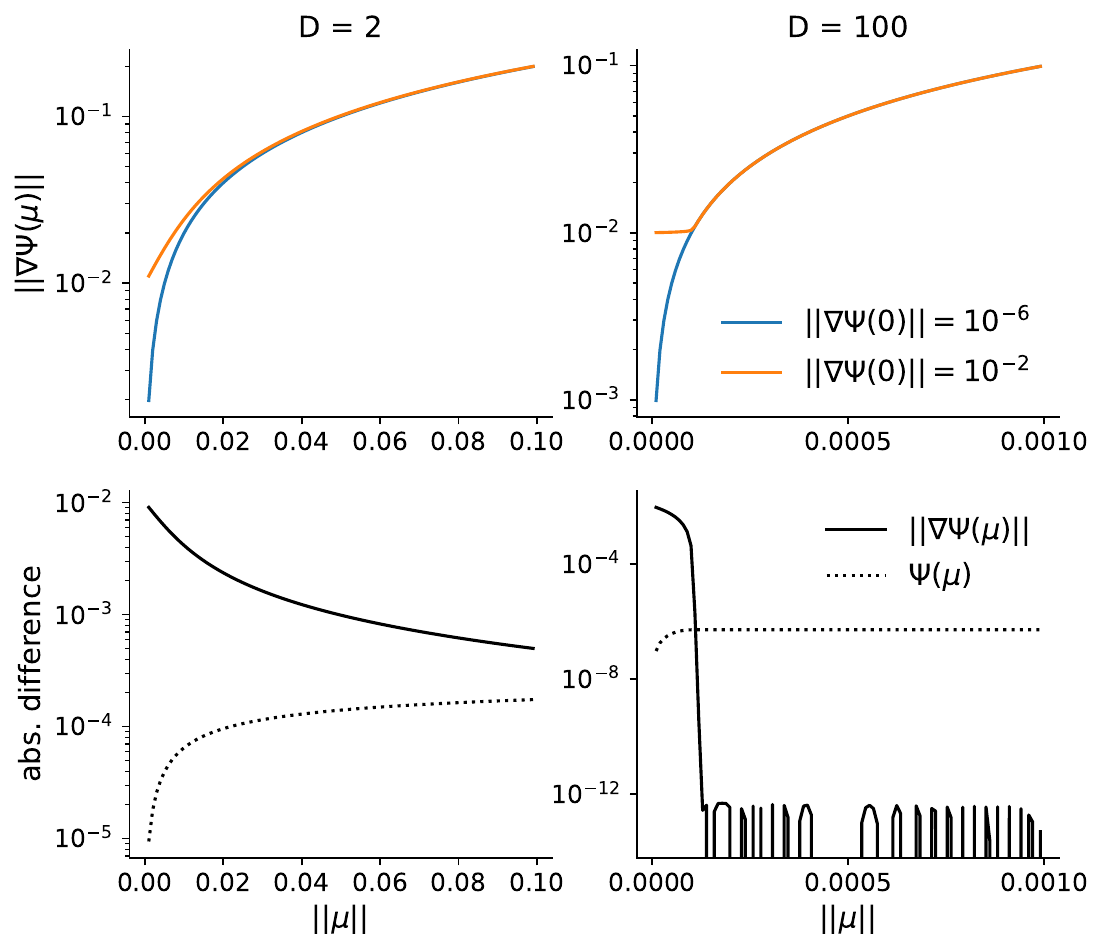}
    \vspace{-0.8cm}
    \caption{Dependence of ODE solutions on initial condition. Top: Numerical solutions $||\grad\Psi(\mu)||=\psi'(\mu)$ to second-order ODE solved from $\psi'(0) = 10^{-6}$ (blue) and from $\psi'(0) = 10^{-2}$ (orange). Solutions for first derivative $\psi'$ approach one another quickly, in particular for larger dimension $D$. Bottom: resulting absolute difference between ODE solutions shown at the top, for radial profile of the negative entropy (dotted line), and the first derivative (solid line).  }
    \label{fig:figS1}
\end{figure}

Solving the ODE for the radial profiles $\psi(||\mu||), \psi'(||\mu||)$ from the center $||\mu||=0$ outwards is problematic due to the initial condition $\psi'(0) = 0$, which causes $\psi''(0)$ to become a fraction of zero over zero. We instead initialize $\psi'(0) > 0$ as a small value, and explore the dependence on $\psi'(0)$ in Fig. \ref{fig:figS1} for dimensions $D=2$ and $D=100$. In particular for larger $D$, the effects of different choices for $\psi'(0)$ on the full radial profile $\Psi(\mu)=\psi(||\mu||)$ are limited, because the first derivative converges quickly even from initializations that differ by four orders of magnitude.
This attracting behavior of the solution for $||\grad\Psi(\mu)||=\psi'(||\mu||)$ in the forward direction also makes it very difficult to solve the ODE in reverse, e.g. from $||\mu|| =0.5$ towards $||\mu||=0$.
For numerical results in this study, we initialized the ODE at values between $\psi'(0)=10^{-6}$ ($D=2$) and $\psi'(0) = 10^{-4}$ ($D=10000$).

\subsection{Closed-form approximation of negative entropy}
\label{appendix:diff_ODE}

We used the following closed-form approximation to the negative entropy of the vMF distribution in $D$ dimensions, along with its first two derivatives:
\begin{align}
    \tilde{\psi}_1(||\mu||) &= \frac{(1-D)}{2} \log(1-||\mu||^2) \nonumber \\  &+ \frac{1-D}{4s}\left(\log(v + ||\mu||^2 + s)- \log(v + s)\right)  \nonumber \\ &- \frac{1-D}{4s}\left(\log(v + ||\mu||^2 - s)- \log(v - s)\right), \nonumber \\
    \tilde{\psi}_1'(||\mu||) 
    &= \frac{(D-1)||\mu||}{1-||\mu||^2} + \frac{(D-1)||\mu||}{||\mu||^4 + (D-2) ||\mu||^2 + D-1},  \nonumber  \\
    \tilde{\psi}_1''(||\mu||) &= (D-1) \frac{1+||\mu||^2}{(1-||\mu||^2)^2} \nonumber \\ 
    &+ (D-1) \frac{- 3||\mu||^4 - (D-2) ||\mu||^2 + D - 1}{(||\mu||^4 + (D-2) ||\mu||^2 + D-1)^2}, \nonumber
\end{align}
where $v =\frac{D}{2}-1$ and $s = \sqrt{v^2 - D}$.
We included the second derivative here since it is needed to compute the vMF variance function. 

We note that further refinements to the approximation to $\psi'$ are possible through another iteration 
\begin{align}
    \tilde{\psi}'_2(||\mu||) &=  \frac{(D-1) ||\mu||}{1- ||\mu||^2 - 1/\tilde{\psi}_1''(||\mu||) }, \nonumber \\
&=  \frac{(D-1)||\mu||}{1-||\mu||^2} \nonumber \\&+ \frac{(D-1) ||\mu|| \alpha(||\mu||)^2}{(D(1+||\mu||^2)-2)\alpha(||\mu||)^2 + \beta(||\mu||)}, \nonumber
\end{align}
where $\alpha(||\mu||) = (D-1)(1+||\mu||^2) - ||\mu||^2(1-||\mu||^2)$, $\beta(||\mu||) = (1-||\mu||^2)^2 (-3||\mu||^4 + (2-D) ||\mu||^2 +D-1)$.
Beyond that, we found the next iteration $\tilde{\psi}_3'$ to do worse than $\tilde{\psi}_2'$ and fall between $\tilde{\psi}_1'$ and $\tilde{\psi}_2'$ in terms of approximation quality. Since the antiderivative $\tilde{\psi}_2$ of $\tilde{\psi}_2'$ involves solving a fifth-order polynomial in $D$, we approximated the negative entropy with the simpler $\tilde{\psi}_1(||\mu||) \approx \Psi(\mu)$ given above.

The integral representations of 
\begin{align}
    \tilde{\psi}(||\mu||) &=\int_0^{||\mu||}\tilde{\psi}'(r) dr, \nonumber \\
    \tilde{\psi}'(||\mu||) &= \int_0^{||\mu||}\tilde{\psi}''(r) \nonumber
\end{align}
are analogous to those of $\psi$ and $\psi'$, allowing to express 
\begin{align}
    \psi(||\mu||) &= \tilde{\psi}(||\mu||) + f(||\mu||),\nonumber \\
    \psi'(||\mu||) &= \tilde{\psi}'(||\mu||) + f'(||\mu||), \nonumber
\end{align}
where $(f, f')$ is the solution to the second-order ODE arising from 
\[ f'' = \frac{\tilde{\psi}'(t) + f'}{(1-t^2) (\tilde{\psi}'(t) + f')) + (1-D) t} - \tilde{\psi}''(t).\]
For large $D$, we found the correction terms $f(||\mu||), f'(||\mu||)$ to become increasingly irrelevant, so we can try solving the ODE with lower precision, or omit it entirely ($f = f' = 0$).

\subsection{Numerical comparison of gradient approximation.}
\label{appendix:newton}

\begin{figure}[h!]
    \centering
    \includegraphics[width=0.50\textwidth]{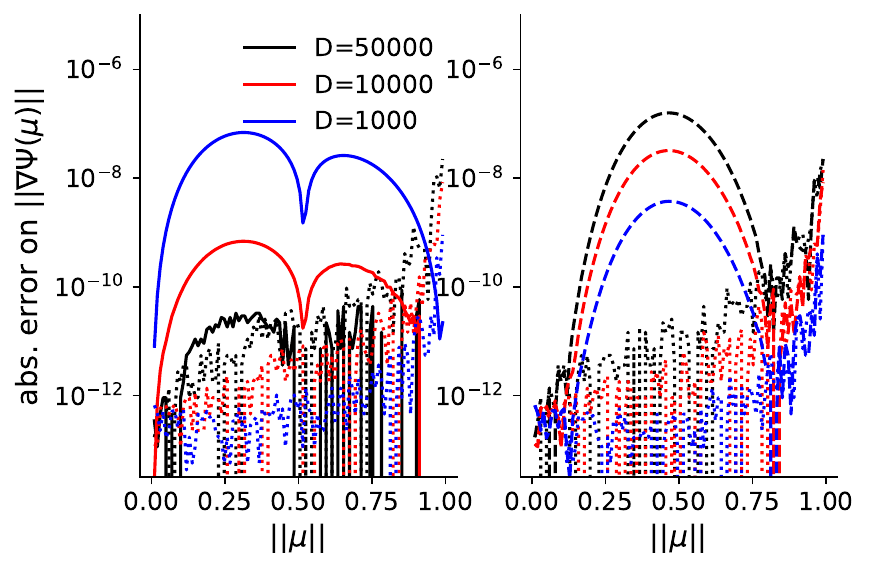}
    \vspace{-0.8cm}
    \caption{Direct comparison of the approximation $\tilde{\psi}_2'(||\mu||) \approx ||\nabla\Psi(\mu)|| = \kappa$ against truncated Newton-Raphson for finding the root of $\phi'(\kappa) - ||\mu||$. Left: Absolute errors $|\hat{\kappa} - ||\nabla\psi(\mu)|| \|$ with $||\nabla\psi(\mu)||= \psi'(||\mu||)$ from the numerical solution of the second-order ODE and $\hat{\kappa}$ estimated via $\tilde{\psi}_2'$ (solid lines) and Newton-Raphson (dotted lines) with $M=2$ steps and $L=20$ iterations to estimate $\phi'(\kappa)$. Right: Same as left, but for $\hat{\kappa}$ being estimated with Newton-Raphson with either more (dotted, $M=3$, $L=50$) or  less computations (dashed, $M=2$, $L=15$). At $D=50.000$ (black), the tenth-degree rational function $\tilde{\psi}_2$ performs at around numerical precision (double-precision floating points), and thus comparable to the more expensive iterative root finder.} 
    \label{fig:figS2}
\end{figure}

We directly compared the numerical approximation $\tilde{\psi}_2'(\mu) \approx ||\nabla\Psi(\mu)||=\kappa$ against root-fining on $\phi(\kappa) - \mu$, as is commonly used in maximum likelihood. We compared against truncated Newton-Raphson \cite{banerjee2005clustering, sra2012short}, which requires a numerical evaluation of $\phi'(\kappa) = \frac{I_{D/2}(\kappa)}{I_{D/2-1}(\kappa)}$, which in turn is non-trivial for large $D$. We followed Hornik and Gr\"{u}n \yrcite{hornik2014maximum} in truncating a Perron continuous fraction representation \cite{gautschi1978computation} of 
\[\frac{I_{v}(\kappa)}{I_{v-1}(\kappa)} = \frac{\kappa}{2v + \kappa - \frac{(2v+1)\kappa}{2v + 1 + 2\kappa - \frac{(2v+3)\kappa}{2v + 2 + 2\kappa - \ldots}}}.\] 
This root finding scheme thus includes two nested loops: an outer loop over $M$ Newton-Raphson steps, and an inner loop over $L$ iterations of the continued fraction. The initialization was given by $\psi_B'(||\mu||)$ \cite{banerjee2005clusteringhyper}.
We found empirically that for $D>10$ truncating the infinite continuous fraction after about $L=20$ iterations gave good results up to numerical precision of double-precision floating point numbers. Every such iteration beyond the first two requires six algebraic operations. We could alternatively have used a Gauss continuous fraction representation instead, which would incur five algebraic operations per iterations but be less stable \cite{hornik2014maximum}. At double precision, convergence typically required $M=2$ or $M=3$ Newton-Raphson steps, for several hundred algebraic operations overall. Evaluating $\tilde{\psi}_2'(||\mu||)$ for given $D, ||\mu||$ in comparison required $23$ algebraic operations or less. We directly compared the two methods of estimating $\kappa = \psi'(||\mu||)$ for large $D$ in figure \ref{fig:figS2}.

\subsection{Datasets and pre-processing}
\label{sec:datasets}
We considered the 'classic3' and 'news20' datasets previously used in the natural language processing literature \cite{mitchell1997machine, bisson2008chi, wang2016structural}. \\ 'classic3' comprises the $K_{true}=3$ document collections CRANFIELD ($N_{\tiny{}CRAN}=1400$), CISI ($N_{\tiny{}CISI}=1460$) and MEDLINE ($N_{\tiny{}MED}=1033$) of abstracts from scientific publications in engineering, information science and medical journals, respectively. The goal of clustering the documents is to assign the abstracts to their respective scientific field. \\
'20 newsgroups' ('news20') is a collection of $19997$ messages from users on USENET newsgroups, distinguished by $K_{true}=20$ different subject matters including religion, computer graphics and motorcycles.  
Removing duplicates, we were left with $N=18803$ individual messages. We removed message headers to avoid the possibility of trivial assignment of messages to subject matter. The goal of clustering was to assign each document to its subject matter. \\
We also followed previous work \cite{banerjee2005clusteringhyper} in creating smaller versions of these datasets by sampling $100$ documents per category from both 'classic3' and 'news20', resulting in the much smaller 'classic300' and 'news20-small' datasets with perfectly balanced labels. 

For each of these datasets, we created a feature mapping from documents onto feature representations on the unit hypersphere with term-frequency inverse term frequency, which starts out from the occurrence counts $v_d^n$ of feature $d=1,\ldots,D$ in document $n=1,\ldots,N$. These term frequencies were weighted by the (log-) inverse document frequencies $g_d = \log(N/\sum_n v_d^n)$ and subsequently normalized per document to account for varying document lengths:
\[x^n = \frac{v^n \odot g}{|| v^n \odot g ||}, \]
where $\odot$ refers to the Hadamard product.

The features indexed $d=1,\ldots,D$ were tokens from a dataset-specific vocabulary. Vocabularies were built from all tokens occurring within a given dataset of $N$ documents, which occurred in at least $P$ different documents ($P$=$7/6/2/2$ for 'news20'/'classic3'/'news20-small'/'classic300', respectively). We chose $P$ to obtain vocabulary sizes $D$ comparable to previous studies. We also removed overly common words appearing in more than $15\%$ of documents, and removed any token that also appears on the SMART list \cite{salton1971smart} of common stopwords (see e.g. appendix 11 of Lewis et al. \yrcite{lewis2004rcv1}).  Documents that did not contain a single token from the final vocabulary ($||v^n \odot g||=0$) were removed. This only happened for two documents from the CRANFIELD collection in the 'classic3' dataset.   

\end{document}